\documentclass[preprint,showpacs,preprintnumbers,amsmath,amssymb,prb]{revtex4-1}

\usepackage[dvipdfmx]{graphicx}
\usepackage{dcolumn}% Align table columns on decimal point
\usepackage{bm}% bold math
\usepackage{mhchem} % chemical molecular formulae and equations
\usepackage{ascmac}
\usepackage{color} %If we use color, figure can be not placed properly. 
\usepackage[normalem]{ulem}
\begin{document}
%\preprint{Submitted to }
%\thispagestyle{empty}
%%%%%%%%%%%%%%%%%%%%%%
%\draft
%\preprint{submitted to ...}
%%%%%%%%%%%%%%%%%%%%%%%%%%%%%%%%%%%%%%%%%%%%%%%%%%%%%%%%%%%%%%%%%%%%%%
% TITLE  %%%%%%%%%%%%%%%%%%%%%%%%%%%%%%%%%%%%%%%%%%%%%%%%%%%%%%%%%%%%%
%Photoluminescence decay quenched by particle surfaces: theoretical study
%%%%%%%%%%%%%%%%%%%%%%%%%%%%%%%%%%%%%%%%%%%%%%%%%%%%%%%%%%%%%%%%%%%%%%
%
\title{Temperature-Gradient Effects on Electric Double Layer Screening in Electrolytes}
\author{
Kazuhiko Seki
}
\email{k-seki@aist.go.jp}
\affiliation{GZR, National Institute of Advanced Industrial Science and Technology (AIST), Onogawa 16-1 AIST West, Ibaraki, 305-8569, Japan
}
%\author{}
%\affiliation{
%}
\date{\today}
%\preprint{}
%%%%%%%%%%%%%%%%%%%%%%%%%%%%%%%%%%%%%%%%%%%%%%%%%%%%%%%%%%%%%%%%%%%%%%
% ABSTRACT  %%%%%%%%%%%%%%%%%%%%%%%%%%%%%%%%%%%%%%%%%%%%%%%%%%%%%%%%%%
%%%%%%%%%%%%%%%%%%%%%%%%%%%%%%%%%%%%%%%%%%%%%%%%%%%%%%%%%%%%%%%%%%%%%%
%\thispagestyle{empty}
%\begin{center}
\begin{abstract}
Temperature gradients drive asymmetric ion distributions via thermodiffusion (the Soret effect), 
leading to deviations from the classical Debye--H\"uckel potential.
We introduce the Eastman entropy of transfer, $\hat{S}_\pm = \alpha_\pm k_{\rm B}$ for cations and anions, 
respectively, where $k_{\rm B}$ is the Boltzmann constant, and analyze non-isothermal electric double layers 
in terms of the dimensionless Soret coefficients $\alpha_\pm$.  
Analytical solutions of the generalized Debye--H\"uckel equation show that, for $\alpha_+ = \alpha_-$, 
the potential is exactly described by a modified Bessel function, while the marginal case $\alpha_\pm = 1$ 
exhibits algebraic decay.  
An effective screening length, $\lambda_{\rm eff}$, characterizes the near-electrode potential and increases 
with temperature, resulting in weaker screening on the hot side and stronger screening on the cold side for 
$\alpha_\pm > -1$.  
The differential capacitance is controlled by $\alpha_\pm$ via $\lambda_{\rm eff}$, with its minimum coinciding 
with the potential of zero charge (PZC) even in the presence of a temperature gradient.  
These findings highlight the fundamental coupling between electrostatics and thermodiffusion in 
non-isothermal electrolytes.
\end{abstract}
%\PACS 78.55.Qr \sep 72.20.Ee \sep  05.40.a

\maketitle
%\newpage
%%%%%%%%%%%%%%%%%%%%%%%%%%%%%%%%%%%%%%%%%%%%%%%%%%%%%%%%%%%%%%%%%%%%%%
% Introduction %%%%%%%%%%%%%%%%%%%%%%%%%%%%%%%%%%%%%%%%%%%%%%%%%%%%%%%%%%%%%%
%%%%%%%%%%%%%%%%%%%%%%%%%%%%%%%%%%%%%%%%%%%%%%%%%%%%%%%%%%%%%%%%%%%%%%
%\setcounter{equation}{0}
\section{Introduction}

Energy harvesting from temperature gradients is a central challenge in sustainable technologies, and ionic thermoelectrics have emerged as a promising class of materials for utilizing low-grade waste heat.%
\cite{Zhao_16,Al-zubaidi_17,Horike_20,Horike_21JMCS,ZHAO_2021,Mardi_21,Mardi_22,Hanlin_22,Zhou_22,Song_22,Sun_23,Horike_24_AMT,Qian_24,JIA_24,Jabeen_24,HORIKE_25CEJ,Yu-Syuan_25,Ates_25,WANG_26} 
According to a recent review,%
\cite{Sun_23,Jabeen_24}
ionic thermoelectrics can achieve thermoelectric potentials in the range of $1$--$10$~mV~K$^{-1}$, 
which is several orders of magnitude larger than the Seebeck coefficients of typical electronic thermoelectrics.  
Such large thermoelectric responses likely originate from the coupling between ionic transport and thermal fields, 
making ionic thermoelectrics attractive candidates for heat-to-electricity energy conversion applications.

In ionic thermoelectric systems, charge carriers are redox-free ions.  
Since the ionic valencies remain unchanged, the total numbers of cations and anions are conserved.  
Nevertheless, local charge neutrality can be violated.  
It has been proposed that differences in ionic polarization near the hot and cold electrodes generate an open-circuit voltage,%
\cite{Horike_20,Horike_21JMCS,Hanlin_22,Al-zubaidi_17,Lim_15,Mardi_21}
leading to the observed thermoelectric potential.  
Unlike electronic thermoelectrics, charge flow to the external circuit is suppressed during the charging process;  
the energy conversion mechanism is therefore capacitive in nature and fundamentally distinct from that of electronic thermoelectrics.

Despite this promising phenomenology, 
the microscopic electrostatic response of ions under non-isothermal conditions remains poorly understood.  
In conventional isothermal electrolyte theory, the Debye--H\"uckel approximation to the Poisson--Boltzmann equation describes an exponential decay of the electrostatic potential, characterized by the Debye screening length.
However, when thermodiffusion (the Soret effect) acts concurrently with electrostatic screening, 
the ionic distributions become dependent on the temperature gradient, 
and it remains unclear whether the classical exponential potential profile is still valid under non-isothermal conditions.

Recently, 
electrostatic screening in ionic fluids subject to stationary temperature gradients has been investigated 
using the hypernetted-chain (HNC) approximation within density functional theory, assuming local equilibrium.%
\cite{Grisafi_23}
Those results indicate that a temperature-gradient can influence the transition between monotonic and oscillatory screening behaviors,  
and their asymptotic analysis suggests deviations from equilibrium screening.  
This finding implies that screening itself, not merely ionic currents, can be fundamentally modified by temperature gradients.

In this work, 
we develop a complementary theoretical framework based on transport equations that explicitly incorporate thermodiffusion through 
the Eastman entropy of transfer, $\hat{S}_\pm = \alpha_\pm k_{\rm B}$ for cations and anions, 
respectively, where $k_{\rm B}$ is the Boltzmann constant. 
We study non-isothermal electric double layers 
in terms of the dimensionless Soret coefficients $\alpha_\pm$.  
In our previous studies,%
\cite{HORIKE_25}
we investigated the Soret effect in electrolytes and the associated thermal ion adsorption/desorption at electrodes as the origin of polarized states, 
showing that thermoelectric potentials on the order of millivolts per kelvin can be generated.  
We also derived a generalized Grahame equation from a generalized Poisson--Boltzmann model without linearization 
to explain the large thermoelectric potentials observed in ionic thermoelectrics.  

Here, we linearize the generalized Poisson--Boltzmann equation to derive a modified 
Debye--H\"uckel equation under a linear temperature gradient. 
Exact analytical solutions are obtained for symmetric cases, and approximate forms are developed for more general conditions, 
leading to the definition of an effective screening length, $\lambda_{\rm eff}$.  
We further analyze how thermodiffusion affects the differential capacitance and the potential of zero charge (PZC). 
Our general theoretical results justify our previous numerically based conclusion that the 
Stern-layer capacitance is the dominant source of the large Seebeck coefficients observed 
in ionic thermoelectric supercapacitors.\cite{HORIKE_25}

Understanding how the Soret effect modifies the electric double layer under temperature gradients 
is crucial not only for ionic thermoelectrics but also for a variety of colloidal and electrokinetic phenomena.%
\cite{Dhont_08,Sarkar_19,Dhont_25}
With this broader perspective, we also extend our discussion to electric double layers around both planar and spherical geometries.

%%%%%%%%%%%%%%%%%%%%%%%%%%%%%%%%%%%%%%%%%%%%%%%%%%%%%%%%%%%%%%%%%%%%%%%
\section{Theoretical formulation}
\label{sec:Theory}

The local concentrations of mobile cations and anions at position $\vec{r}$ are denoted by $n_+(\vec{r})$ and $n_-(\vec{r})$, respectively. 
Here, $+$ and $-$ specify cationic and anionic properties, respectively. 
The ionic valencies are denoted by $z_\pm$, and $q$ denotes the elementary charge. 
For instance, monovalent cations and anions correspond to $z_+=1$ and $z_-=-1$, respectively. 
The diffusion constants of cations and anions are denoted by $D_+$ and $D_-$, respectively. 
According to the Einstein relation, the diffusion constant is related to the corresponding mobility as 
\begin{align}
D_\pm(\vec{r})=\mu_\pm k_{\rm B} T(\vec{r}),
\end{align}
where $k_{\rm B}$ is the Boltzmann constant, $T(\vec{r})$ is the local temperature, and isotropic diffusion is assumed. 

In free space without boundaries, a temperature gradient induces a mass current density proportional to the Eastman entropy of transfer.  
We consider the case where the Eastman entropy of transfer, $\hat{S}_\pm$, is given by $\hat{S}_\pm = \alpha_\pm k_{\rm B}$, \cite{Eastman_28,Wurger_08} 
with $\alpha_\pm$ denoting dimensionless Soret coefficients.  
The Eastman entropy of transfer scales with $k_{\rm B}$, consistent with the classical expression derived from the Boltzmann equation. \cite{Mahan_98}  
For point-like ions in a spatially varying temperature field, $\alpha_\pm = 1$ has been derived from the Fokker--Planck equation. \cite{vanKampen_88}  
According to a theory based on the Born solvation model, the dimensionless Soret coefficient 
$\alpha_{\pm}$ is, in its simplest form, predicted to be independent of temperature.\cite{Agar_89} 
Experimentally, the Eastman entropy of transfer is not measured directly but is instead inferred 
from Soret data.\cite{Sugisaki_06,Caldwell_81} In such measurements, the Soret coefficient 
of a neutral electrolyte does not probe cationic and anionic contributions separately, but rather a 
specific linear combination of the underlying ionic entropies of transport. At the salt level, 
experimental determinations of Soret coefficients and transport entropies in aqueous electrolytes 
are consistent with only weak temperature dependence over modest intervals (on the order of 
$\sim 10$~K) around room temperature.\cite{Sugisaki_06,Caldwell_81}
Although Soret coefficients have been measured for both ions and macroparticles, \cite{Kishikawa_10,CabreiraGomes_18,Sehnem_18}  
their general temperature dependence remains unclear.  
Here, we assume that the dimensionless coefficients are temperature independent, as in the case of point particles ($\alpha_\pm = 1$).  
The probability current density $\vec{j}_{\pm}$ is then given by \cite{Wurger_08,Bonetti_11,Hanlin_22,Wurger_10,Stout_17,KIM_18,ZHANG_19,JIA_24,Jabeen_24,HORIKE_25}
\begin{align}
\vec{j}_{\pm} 
&= - D_{\pm} \left( 
\mathrm{grad}\, n_{\pm} - \frac{z_\pm q \vec{E} n_{\pm}}{k_{\rm B} T }
+\frac{\hat{S}_\pm}{k_{\rm B} T } n_{\pm}\,\mathrm{grad}\, T
\right) ,
\label{eq:jpm0_s}
\end{align}
where $\vec{E}$ denotes the electric field.
We introduce the electrostatic potential $\psi(\vec{r})$, satisfying 
\begin{align}
\vec{E}(\vec{r}) = - \mathrm{grad}\, \psi(\vec{r}). 
\label{eq:psi}
\end{align}

Under open-circuit conditions, we set $\vec{j}_{\pm} =0$, yielding
\begin{align}
0
&= 
k_{\rm B} T \,\mathrm{grad}\, n_{\pm}
- z_\pm q \vec{E} n_{\pm}
+ k_{\rm B} \alpha_{\pm} n_{\pm}\,\mathrm{grad}\, T .
\label{eq:p1}
\end{align}
Since both $n_{\pm}$ and $T$ vary with position $\vec{r}$, we divide Eq.~(\ref{eq:p1}) by $n_{\pm}k_{\rm B} T$ to obtain
\begin{align}
0
&= 
\frac{1}{n_{\pm}} \,\mathrm{grad}\, n_{\pm}
-\frac{z_\pm q \vec{E}}{k_{\rm B} T}
+ \alpha_{\pm} \frac{1}{T}\,\mathrm{grad}\, T .
\label{eq:p1_r1}
\end{align}
Equation~(\ref{eq:p1_r1}) can be rewritten as
\begin{align}
\mathrm{grad}\, \ln \!\left(n_{\pm} T^{\alpha_{\pm}} \right) 
&= 
 \frac{z_\pm q \vec{E}}{k_{\rm B} T}.
\label{eq:p2_5}
\end{align}

Introducing the radial coordinate $r$ and assuming isotropy, we define the outward normal unit vector as $\vec{n}=\vec{r}/|\vec{r}|$. 
By integration, we obtain the formal solution
\begin{align}
n_\pm(r) 
&= n_\pm(L)\left( \frac{T(L)}{T(r)} \right)^{\alpha_\pm}
\exp\left( -\int_r^L  dr_1 \frac{z_\pm q \vec{E} (r_1)\cdot \vec{n} }{k_{\rm B} T(r_1)} \right) . 
\label{eq:7rw}
\end{align}
From Gauss's law, we have
\begin{align}
\epsilon_{\rm r} \epsilon_0\,\mathrm{div}\, \vec{E}(\vec{r})
= q \sum_{j} z_j n_j(\vec{r}), 
\label{eq:Gauss}
\end{align}
where $n_j (\vec{r})$ denotes $n_+(\vec{r})$ or $n_-(\vec{r})$, $z_j$ denotes $z_+$ or $z_-$, and $z_-=-z_+$ with $z_+>0$. 
The Poisson--Boltzmann equation generalized to account for spatially inhomogeneous temperature distributions is then expressed as \cite{HORIKE_25}
\begin{align}
\nabla^2 \,\psi(r)
= -\frac{q}{\epsilon_{\rm r} \epsilon_0} \sum_{j} z_j 
n_j (L)\left( \frac{T(L)}{T(r)} \right)^{\alpha_j}
\exp \!\left(\int_r^L dr_1 \frac{z_j q}{k_{\rm B} T(r_1)} \frac{d \psi}{d r_1} \right). 
\label{eq:PB}
\end{align}

As shown in Appendix A, for a small temperature gradient, we obtain by linearizing Eq.~(\ref{eq:PB}), 
\begin{align}
\nabla^2 \psi(r) = \frac{q^2}{\epsilon_{\rm r} \epsilon_0 k_{\rm B} T(L)} 
\sum_{j} z_j^2 n_j (L)
\left(\frac{T(L)}{T(r)} \right)^{1+\alpha_j} 
\psi(r),
\label{eq:gGC1}
\end{align}
where we set $\psi(L)=0$, and $\psi(r)$ hereafter denotes $\psi(r)-\psi(L)$. 
%%% cut 

Equation~(\ref{eq:gGC1}) can be recast as
\begin{align}
\nabla^2 \psi(r) 
= \frac{\kappa^2}{\sum_{j} z_j^2} 
\sum_{j} z_j^2 
\left(\frac{T(L)}{T(r)} \right)^{1+\alpha_j} 
\psi(r),
\label{eq:DH}
\end{align}
where $\kappa$ is the Debye--H\"{u}ckel parameter defined by \cite{Debye_23}
\begin{align}
\kappa^2 = \frac{q^2}{\epsilon_{\rm r} \epsilon_0 k_{\rm B} T(L)} 
\sum_{j} z_j^2 n_j (L). 
\end{align}

The characteristic thickness of the electric double layer is given by $1/\kappa$. 
Assuming $L \gg 1/\kappa$ and define the Debye--H\"{u}ckel screening length as $\lambda_{\rm D}=1/\kappa$, which can be expressed as
\begin{align}
\lambda_{\rm D} = \left(\frac{\epsilon_{\rm r} \epsilon_0 k_{\rm B} T(L)}
{2q^2 N_{\rm A}} \right)^{1/2} \frac{1}{\sqrt{I}}, 
\label{eq:DH_1}
\end{align}
where $N_{\rm A}$ is Avogadro’s constant and 
$I$ is the ionic strength, defined by
\begin{align}
I = \frac{1}{2}\sum_{j} z_j^2 \left(\frac{n_j(L)}{N_{\rm A}} \right).
\label{eq:I}
\end{align}
Equations (\ref{eq:DH})--(\ref{eq:I}) provide the basis for further analysis of electric double layer profiles.
%%% English edited

%%%%%%%%%%%%%%%%%%%%%%%%%%%%%%%%%%%%%%%%%%%%%
\section{One Dimension}
%%%%%%%%%%%%%%%%%%%%%%%%%%%%%%%%%%%%%%%%%%%%%

%%%%%%%%%%%%%%%%%%%%%%%%%%%%%%%%%%%%%%%%%%%%%%%%%%%%%%%%%%%%%%%%%%%%%%%
\begin{figure}[h]
\begin{center}
\includegraphics[width=0.5\textwidth]{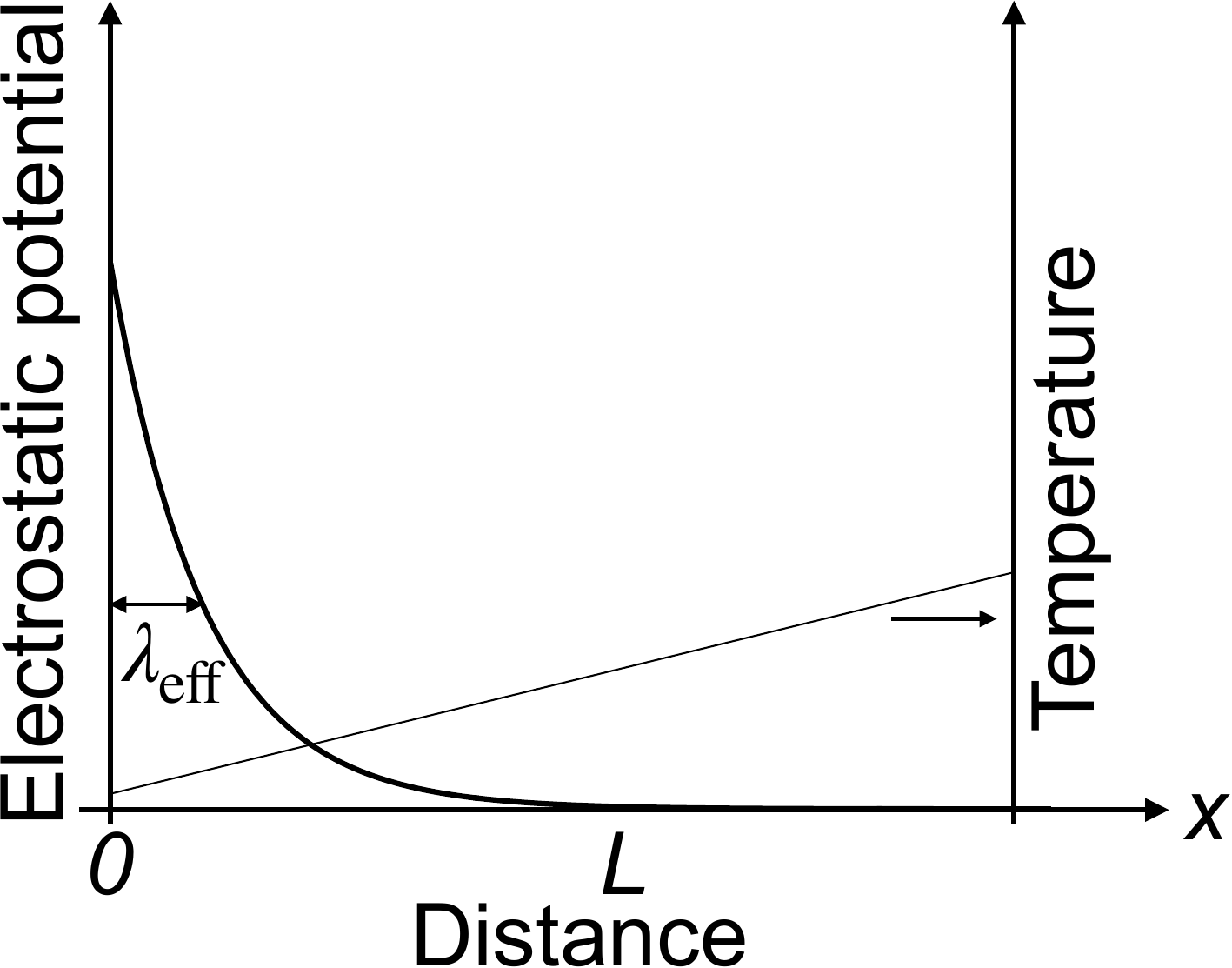}
\end{center}
\caption{%(Color online) 
Schematic of the electrostatic potential in a linear temperature gradient. 
The origin is located at the edge of the diffuse layer, and charge neutrality is satisfied at the distance $L$. 
The parameter $\lambda_{\rm eff}$ denotes the effective screening length.
}
\label{fig:schematicfig}
\end{figure}
%%%%%%%%%%%%%%%%%%%%%%%%%%%%%%%%%%%%%%%%%%%%%%%%%%%%%%%%%%%%%%%%%%%%%%%

We introduce a one-dimensional coordinate to describe the positions of mobile ions in the electrolyte 
as shown in Fig. \ref{fig:schematicfig}. 
The $x$ axis is defined normal to the electrode, with the origin at the electrode surface and $x$ increasing with distance from the electrode. 
We assume that the charge neutrality condition is satisfied at $x=L$.   

In one dimension, Eq.~(\ref{eq:DH}) can be rewritten as
\begin{align}
\frac{\partial^2}{\partial x^2} \psi(x)
= \frac{\kappa^2}{\sum_{j} z_j^2 } \sum_{j} z_j^2 
\left(\frac{T(L)}{T(x)}\right)^{1+\alpha_j} \psi(x).
\label{eq:DH1dim}
\end{align}
Previously, Eq.~(\ref{eq:DH1dim}) with $\alpha_\pm = 1$ was extended to include the temperature dependence of the dielectric constant and solved perturbatively. \cite{Dhont_08}  
For simplicity, we neglect the spatial variation of the dielectric constant arising from its temperature dependence.

We assume a linear temperature profile between $T_0=T(0)$ and $T(L)$, \cite{Stout_17,Jabeen_24,HORIKE_25}
\begin{align}
\frac{T(x)}{T(L)}=T_{\rm r}+g x, 
\label{eq:Tx}
\end{align}
where $T_{\rm r}=T_0/T(L)$ indicates the reference temperature, and the gradient is defined as
\begin{align}
g=\frac{1-T_{\rm r}}{L}.
\label{eq:g}
\end{align}

%%%%%%%%%%%%%%%%%%%%%%%%%%%%%%%%%%%%%%%
\subsection{Approximate Solution}
%%%%%%%%%%%%%%%%%%%%%%%%%%%%%%%%%%%%%%%

We next derive approximate analytical forms of the potential under a temperature gradient.  
As shown in Appendix B, 
an approximate solution of Eq.~(\ref{eq:DH1dim}) can be obtained as 
\begin{align}
\psi(x)\approx \psi_{\rm c} \left(\sum_{j} z_j^2 
\left(\frac{1}{T_{\rm r}+g x}\right)^{1+\alpha_j}\right)^{-1/4} 
\exp \left(-x/\lambda_{\rm eff} \right),
\label{eq:asym3}
\end{align}
where an effective screening length under a temperature gradient is defined as
\begin{align}
\lambda_{\rm eff} 
=\lambda_{\rm D} \frac{\sqrt{\sum_{j} z_j^2}}{\sqrt{\sum_{j} z_j^2 
\left(1/T_{\rm r}\right)^{1+\alpha_j}}} ,
\label{eq:lmdeff}
\end{align}
and $\lambda_{\rm D}$ is the Debye--H\"{u}ckel length under isothermal conditions [Eq.~(\ref{eq:DH_1})].
Equation (\ref{eq:lmdeff}) is derived by linearizing the 
electrostatic potential difference and is valid for arbitrary combinations of $\alpha_\pm$ within 
this linear-response framework. 

%%%%%%%%%%%%%%%%%%%%%%%%%%%%%%%%%%%%%%%%%%%%%%%%%%%%%%%%%%%%%%%%%%%%%%%
\begin{figure}[h]
\begin{center}
\includegraphics[width=1\textwidth]{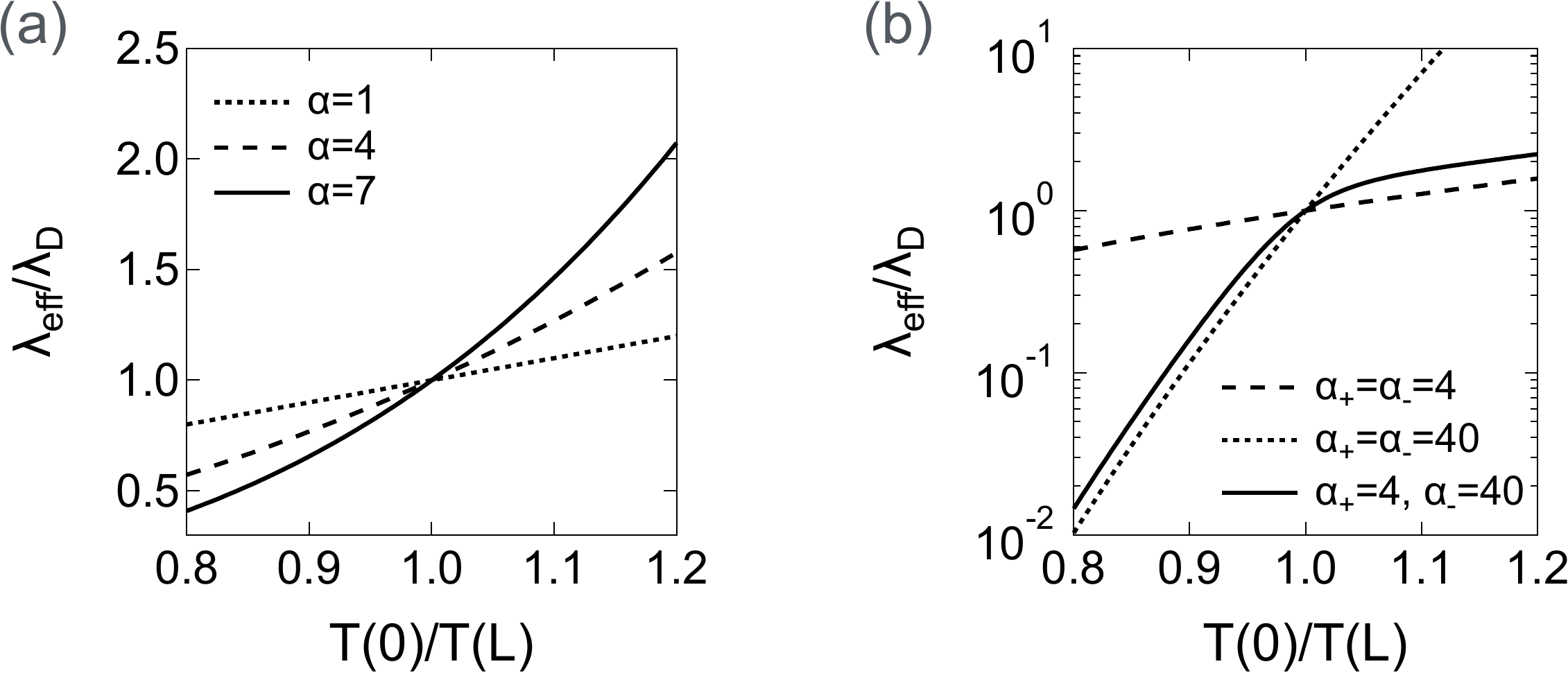}
\end{center}
\caption{%(Color online) 
$\lambda_{\rm eff}$ calculated from Eq.~(\ref{eq:lmdeff}). 
$T(0)$ and $T(L)$ denote the local temperature inside the electric double layer and 
the bulk temperature, respectively, as illustrated in Fig.~\ref{fig:schematicfig}. 
(a) Short-dashed, long-dashed, and solid lines correspond to 
$\alpha_+ = \alpha_- = 1$, $\alpha_+ = \alpha_- = 4$, and $\alpha_+ = \alpha_- = 7$, respectively.  
(b) Long-dashed, short-dashed, and solid lines correspond to 
$\alpha_+ = \alpha_- = 4$, $\alpha_+ = \alpha_- = 40$, and $\alpha_+ = 4, \alpha_- = 40$, respectively.
}
\label{fig:lmd}
\end{figure}
%%%%%%%%%%%%%%%%%%%%%%%%%%%%%%%%%%%%%%%%%%%%%%%%%%%%%%%%%%%%%%%%%%%%%%%

When $\alpha_j > 0$, $\lambda_{\rm eff}$ increases (decreases) with increasing (decreasing) $T_{\rm r}$, corresponding to a higher (lower) electrode temperature, as shown in Fig.~\ref{fig:lmd}.  
In the absence of the Soret effect ($\alpha = 0$), Eq.~(\ref{eq:lmdeff}) reduces to the ordinary temperature dependence of the Debye--H\"{u}ckel length near $x = 0$.  
Thermodiffusion (the Soret effect) is incorporated through the parameter $\alpha$, which modifies $\lambda_{\rm eff}$.  
For $\alpha_+ > 0$ and $\alpha_- > 0$, ions accumulate on the cold side, enhancing screening and thereby reducing the screening length, whereas ion depletion on the hot side increases the screening length.  
It should be noted that $\lambda_{\rm eff}$ is larger at the lower electrode temperature even when $\alpha = 0$,  \cite{ZHANG_19}
because ions naturally accumulate on the colder electrode side in the absence of the Soret effect.  
This behavior persists when $\alpha_+ = \alpha_- > -1$.

\medskip

This qualitative picture can be quantified as follows.  
Neglecting the electrostatic potential gradient in Eq.~(\ref{eq:p1}), the one-dimensional continuity equation reduces to
\begin{align}
\frac{\partial}{\partial x} n_{\pm} 
= - \frac{\hat{S}_\pm}{k_{\rm B} T } n_{\pm}\frac{\partial}{\partial x}T(x),   
\label{eq:Soret}
\end{align}
where $\hat{S}_\pm/(k_{\rm B} T) = \alpha_{\pm}/T$ can be interpreted as the differential form of the Soret coefficient.  
Integration of Eq.~(\ref{eq:Soret}) yields
\begin{align}
\frac{n_\pm(x)}{n_\pm(L)}=\left( \frac{T(L)}{T(x)} \right)^{\alpha_\pm}, 
\label{eq:Soretresult}
\end{align}
which replaces Eq.~(\ref{eq:7rw}) when the potential gradient is neglected.  
If we interpret the bulk concentration $n_{\pm}(L)$ in the ionic strength, Eq.~(\ref{eq:I}), 
as the local concentration $n_{\pm}(0)$ ({\it i.e.}, the concentration inside the electric double layer) 
in the presence of a temperature gradient,
then substituting Eq.~(\ref{eq:Soretresult}) into 
the isothermal Debye length $\lambda_{\rm D}$ [Eq.~(\ref{eq:DH_1}) together with Eq.~(\ref{eq:I})] 
reproduces the effective screening length $\lambda_{\rm eff}$ given in Eq.~(\ref{eq:lmdeff}).

The above derivation relies on neglecting the electrostatic potential in Eq.~(\ref{eq:p1}) 
and interpreting the bulk concentration as the local concentration inside the double-layer. 
When the potential term is retained, 
the effective screening length $\lambda_{\rm eff}$ given in Eq.~(\ref{eq:lmdeff}) can be derived 
without any ad hoc replacement of the bulk concentration. 
In this case, the resulting potential profile no longer exhibits an exponential-like decay 
[Eq.~(\ref{eq:asym3})], but instead takes the non-exponential form given in Eq.~(\ref{eq:asym1}).
Near the electrode, this non-exponential solution can be approximated by an 
exponential-like decay with a power-law correction [Eq.~(\ref{eq:asym3})], from which 
$\lambda_{\rm eff}$ is obtained. Although $\lambda_{\rm eff}$ characterizes the decay close 
to the electrode, the overall potential profile remains non-exponential. As shown later, 
the exponential-like approximation [Eq.~(\ref{eq:asym3})] is valid only over a limited range.

\medskip

It is worth noting that the electrostatic potential in Eq.~(\ref{eq:p1}) is related to the ionic Seebeck coefficient.  
Under the charge neutrality condition ($\vec{j}_{\nu}=0$ for each species), Eq.~(\ref{eq:p1}) in one dimension with $\sum_{\nu} z_{\nu} q j_{\nu}/D_{\nu}=0$ gives
\begin{align}
-q  \sum_{\nu} z_{\nu}  \left( 
\frac{z_{\nu} q E n_{\nu}}{k_{\rm B} T }+\frac{\hat{S}_{\nu} }{k_{\rm B} T } n_{\nu} \frac{\partial}{\partial x}T
\right) =0 , 
\label{eq:n6}
\end{align}
which leads to
\begin{align}
E=\frac{\sum_{\nu} z_{\nu}  \hat{S}_{\nu} n_{\nu}}{\sum_{\nu} z_{\nu}^2  q n_{\nu}}  \frac{\partial}{\partial x}T.  
\label{eq:n7}
\end{align}
The ionic Seebeck coefficient is therefore \cite{Wurger_20}
\begin{align}
S_{\rm total}=\frac{\sum_{\nu} z_{\nu} \hat{S}_{\nu} n_{\nu}}{\sum_{\nu} z_{\nu}^2 q n_{\nu}}.  
\label{eq:n8}
\end{align}
For a binary electrolyte with $z_+=z$ and $z_-=-z$, this reduces to \cite{Wurger_08,Bonetti_11,Dietzel_16,Wurger_20,HORIKE_25}
\begin{align}
S_{\rm total}=\frac{\hat{S}_+ - \hat{S}_-}{2zq}.  
\label{eq:n9}
\end{align}
Since the ionic Seebeck coefficient in Eq.~(\ref{eq:n9}) is derived under the charge neutrality condition,  
the primary influence of a temperature gradient on the electric double layer thickness arises from the Soret effect rather than the ionic Seebeck effect.  
It should also be noted that the Seebeck coefficient is conventionally defined under the assumption of charge neutrality; \cite{Kjelstrup_23}  
in contrast, the present study investigates the thermoelectric potential profile within the electric double layer, where charge neutrality is locally violated.  
\medskip

When $\alpha_- \neq \alpha_+$ and $\alpha_\pm \neq 1$, 
a more accurate evaluation of Eq.~(\ref{eq:asym1}), compared to the exponential-like approximation 
in Eq.~(\ref{eq:asym3}), can be obtained using a hypergeometric-function representation, as shown in Appendix~B. 
As will be shown later, an exact analytical solution of Eq.~(\ref{eq:p1}) exists only when $\alpha_+ = \alpha_-$.  
In this context, the hypergeometric form serves as a useful guide for assessing the validity of the 
exponential-like potential profile in Eq.~(\ref{eq:asym3}) when $\alpha_+ \neq \alpha_-$ and $\alpha_\pm \neq 1$.

%%%%%%%%%%%%%%%%%%%%%%%%%%%%%%%%%%%%%%%%%%%%%%%
\section{Exact solution for $\alpha_+ = \alpha_- $}
%%%%%%%%%%%%%%%%%%%%%%%%%%%%%%%%%%%%%%%%%%%%%%%

An exact analytical solution can be obtained when the thermodiffusion parameters of cations and anions are identical, i.e., $\alpha_+ = \alpha_- = \alpha$.  
Under this condition, the coupled equations simplify considerably, and Eq.~(\ref{eq:DH}) combined with Eq.~(\ref{eq:Tx}) reduces to
\begin{align}
\frac{d^2\psi(x)}{dx^2} = \kappa^2 
\left(\frac{1}{T_{\rm r}+g x}\right)^{1+\alpha} \psi(x),
\label{eq:gGCa}
\end{align}
where $T_{\rm r}$ is the reference temperature and $g$ denotes the temperature gradient.
As shown in Appendix C, 
the general solution of Eq.~(\ref{eq:gGCa}) can be expressed in terms of modified Bessel functions \cite{zaitsev02,Beals_2010}:
\begin{multline}
\psi(x) = (T_{\rm r}+gx)^{1/2} \Bigg[
C_1 I_{1/|\alpha-1|}\!\left(\frac{2\kappa(T_{\rm r}+gx)^{(1-\alpha)/2}}{g|\alpha-1|}\right)
\\
+\, C_2 K_{1/|\alpha-1|}\!\left(\frac{2\kappa(T_{\rm r}+gx)^{(1-\alpha)/2}}{g|\alpha-1|}\right)
\Bigg],
\label{eq:sol1}
\end{multline}
where the absolute value sign is absent in $(T_{\rm r}+gx)^{(1-\alpha)/2}$. 

\medskip

\textbf{Case 1: $\alpha>1$.}  
For $\alpha>1$, the physically relevant branch of Eq. (\ref{eq:sol1}) is
\begin{align}
\psi(x) = \psi_{\rm c}(T_{\rm r}+gx)^{1/2}
I_{1/(\alpha-1)}\!\left[\frac{2\kappa(T_{\rm r}+gx)^{(1-\alpha)/2}}{g(\alpha-1)}\right], 
\label{eq:sol1_1}
\end{align}
by considering that the electrostatic potential must decay away from the electrode. 

\medskip

\textbf{Case 2: $\alpha=1$.}  
When $\alpha=1$, Eq.~(\ref{eq:gGCa}) becomes an Euler-type equation upon substituting $T_{\rm r}+gx$ as the new variable in place of $x$, leading to \cite{zaitsev02}
\begin{align}
\psi(x)=C_1 (T_{\rm r}+g x)^{\frac{1+\sqrt{1+4\kappa^2/g^2}}{2}}
+ C_2 (T_{\rm r}+g x)^{\frac{1-\sqrt{1+4\kappa^2/g^2}}{2}}.
\label{eq:sol2}
\end{align}
Selecting the decaying branch yields
\begin{align}
\psi(x)=\psi_{\rm c}(T_{\rm r}+g x)^{\frac{1-\sqrt{1+4\kappa^2/g^2}}{2}}.
\label{eq:sol2_1}
\end{align}
This expression explicitly shows that the potential decreases algebraically, rather than exponentially, in the presence of a temperature gradient.

\textbf{Case 3: $\alpha < 1$.}  
For $\alpha<1$, the physically relevant branch of Eq. (\ref{eq:sol1}) is
\begin{align}
\psi(x) = \psi_{\rm c}(T_{\rm r}+gx)^{1/2}
K_{1/(1-\alpha)}\!\left[\frac{2\kappa(T_{\rm r}+gx)^{(1-\alpha)/2}}{g(1-\alpha)}\right], 
\label{eq:sol2_3}
\end{align}
by considering that the electrostatic potential must decay away from the electrode. 
Equation (\ref{eq:sol2_3}) reduces to the Airy function when $\alpha=-2$. 

\medskip

\textbf{Approximate form.}  
To examine the asymptotic behavior, Eq.~(\ref{eq:asym1}) gives
\begin{align}
\int_0^x \!dx_1\, (T_{\rm r}+gx_1)^{-(1+\alpha)/2} =
\begin{cases}
\dfrac{2}{g(\alpha-1)} \!\left[T_{\rm r}^{(1-\alpha)/2}-(T_{\rm r}+gx)^{(1-\alpha)/2}\right], & \alpha\neq1,\\[2mm]
\dfrac{1}{g}\ln\!\left(1+\dfrac{gx}{T_{\rm r}}\right), & \alpha=1.
\end{cases}
\end{align}
For $\alpha\neq1$, this leads to
\begin{align}
\psi(x)\approx \psi_{\rm c}(T_{\rm r}+gx)^{(1+\alpha)/4}
\exp\!\left[\frac{2\kappa}{g(\alpha-1)}(T_{\rm r}+gx)^{-(\alpha-1)/2}\right],
\label{eq:asymalphaneq1}
\end{align}
which agrees with the asymptotic form of the modified Bessel function $I_\nu(z)$ \cite{abramowitzstegun} in Eq.~(\ref{eq:sol1_1}):
\begin{align}
I_\nu(z)\approx \frac{e^{z}}{\sqrt{2\pi z}}.
\label{eq:asymK}
\end{align}

For a weak temperature gradient ($gx\ll T_{\rm r}$), expanding $(T_{\rm r}+gx)^{-(\alpha-1)/2}$ yields
\begin{align}
\psi(x)\approx \psi_{\rm c1}(T_{\rm r}+gx)^{(1+\alpha)/4}
\exp[-\kappa T_{\rm r}^{-(\alpha+1)/2}x].
\label{eq:asymalphaneq1_a1}
\end{align}
This form resembles the isothermal Debye--H\"uckel result but includes a temperature-dependent correction that modifies the decay rate.

\medskip

\textbf{Effective screening length.}  
By analogy with Eq.~(\ref{eq:DH_1}), the temperature-dependent effective screening length is defined as
\begin{align}
\lambda_{\rm eff} = \lambda_{\rm D} T_{\rm r}^{(\alpha+1)/2},
\label{eq:DYtempga2}
\end{align}
which is consistent with Eq.~(\ref{eq:lmdeff}).  
Hence, $\lambda_{\rm eff}$ increases with temperature for $\alpha>-1$;   
screening becomes weaker (larger $\lambda_{\rm eff}$) on the hot side of the electrolyte and stronger (smaller $\lambda_{\rm eff}$) on the cold side.  
This spatial asymmetry in screening reflects the influence of thermodiffusion on charge redistribution near the electrode.

\medskip

\textbf{Limiting case $\alpha=1$.}  
For $\alpha=1$, Eq.~(\ref{eq:asym1}) gives
\begin{align}
\psi(x)\approx \psi_{\rm c}(T_{\rm r}+gx)^{1/2-\kappa/g},
\label{eq:asymalphaeq1}
\end{align}
which coincides with Eq.~(\ref{eq:sol2_1}) in the limit of small temperature gradients ($g^2\ll4\kappa^2$).  
Using $(T_{\rm r}+gx)^{-\kappa/g}\!\approx T_{\rm r}^{-\kappa/g} e^{-\kappa x/T_{\rm r}}$, we recover the same scaling relation as Eq.~(\ref{eq:DYtempga2}),
\begin{align}
\lambda_{\rm eff} = \lambda_{\rm D} T_{\rm r},
\end{align}
confirming that the effective screening length grows linearly with the reference temperature in the marginal case $\alpha=1$.

\medskip
\textbf{Summary.}  
The analytical solutions demonstrate that, under a temperature gradient, the electrostatic potential deviates from the exponential Debye--H\"uckel form.  
Nevertheless, the near-electrode potential can be approximated by an exponential function with an effective screening length, $\lambda_{\rm eff}$, apart from a power-law
correction term.  
The effective screening length increases with temperature, indicating that higher temperatures systematically reduce electrostatic screening relative to the isothermal case.

%%%%%%%%%%%%%%%%%%%%%%%%%%%%%%%%%%%%%%%
\section{Three-Dimensional Isotropic System}
%%%%%%%%%%%%%%%%%%%%%%%%%%%%%%%%%%%%%%%

In a three-dimensional isotropic electrolyte, the potential satisfies
\begin{align}
\frac{1}{r^2} \frac{d}{dr} \left( r^2 \frac{d}{dr} \psi_3(r) \right)
= \frac{\kappa^2}{\sum_j z_j^2} \sum_j z_j^2 
\left(\frac{T(L)}{T(r)}\right)^{1+\alpha_j} \psi_3(r).
\end{align}
Defining $\varphi(r)=r \psi_3(r)$ reduces the equation to the one-dimensional form [Eq.~(\ref{eq:DH1dim}) with $x \to r$], giving
\begin{align}
\psi_3(r) = \frac{\psi(r)}{r}.
\label{eq:3D}
\end{align}
Equation (\ref{eq:3D}) with Eq. (\ref{eq:sol1_1}) ($\alpha_+=\alpha_->1$) or 
Eq. (\ref{eq:sol2_1}) ($\alpha_+=\alpha_-=1$) constitutes the exact solution. 
For $\alpha_+=\alpha_-=\alpha$, an approximate solution is
\begin{align}
\psi_3(r) \approx \psi_{\rm c1} (T_{\rm r}+g r)^{(1+\alpha)/4} 
\frac{\exp(-r/\lambda_{\rm eff})}{r}, \quad
\lambda_{\rm eff} = \lambda_{\rm D} T_{\rm r}^{(\alpha+1)/2},
\end{align}
where $\lambda_{\rm eff}$ is the effective screening length under a temperature gradient.
An approximate solution including the case of $\alpha_+\neq \alpha_-$ can be obtained by substituting 
Eq. (\ref{eq:asym3}) into Eq. (\ref{eq:3D}). 
For $\alpha_+\neq \alpha_-$, the more accurate approximate solution is given by 
Eq. (\ref{eq:3D}) with Eq. (\ref{eq:hyper1}). 

%%%%%%%%%%%%%%%%%%%%%%%%%%%%%%%%%%%%%%%
\section{Numerical Results}
%%%%%%%%%%%%%%%%%%%%%%%%%%%%%%%%%%%%%%%

%%%%%%%%%%%%%%%%%%%%%%%%%%%%%%%%%%%%%%%%%%%%%%%%%%%%%%%%%%%%%%%%%%%%%%%
\begin{figure}[h]
\begin{center}
\includegraphics[width=1\textwidth]{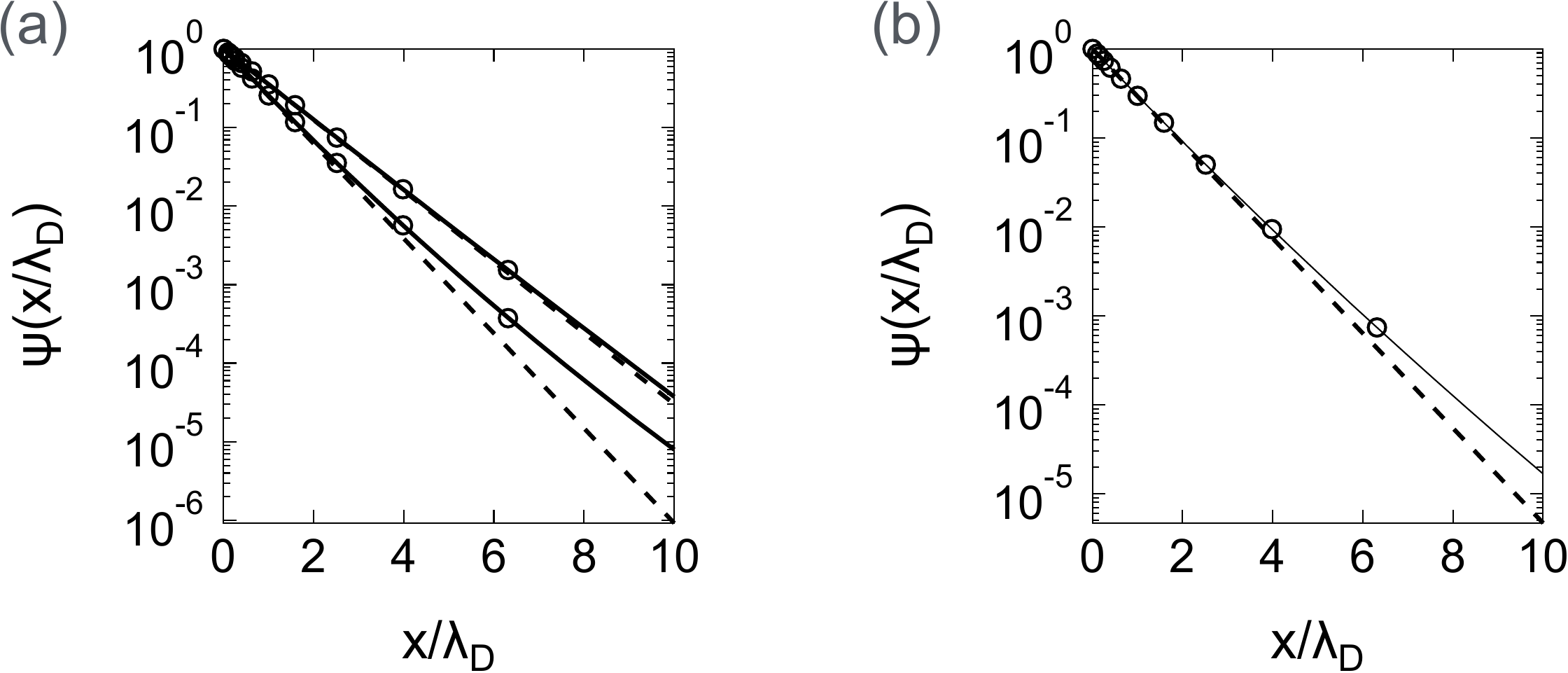}
\end{center}
\caption{Electrostatic potential as a function of the dimensionless distance from the electrode ($x/\lambda_D$) with $\psi(0)=1$, $T_{\rm r}=T(0)/T(L)=0.98$, and $g=0.00165$.  
Thick solid line: exact solution for $\alpha_+=\alpha_-$ [Eq.~(\ref{eq:sol1_1})];  
thin line: hypergeometric approximation [Eq.~(\ref{eq:hyper1})];  
dashed line: exponential decay with a power-law correction [Eq.~(\ref{eq:asym3})] using $\lambda_{\rm eff}$ [Eq.~(\ref{eq:lmdeff})];  
circles: numerical solution of Eq.~(\ref{eq:DH}) with the additional boundary condition $\psi(10\lambda_D)=0$.  
(a) Upper and lower curves/circles correspond to $\alpha_+=\alpha_-=4$ and $\alpha_+=\alpha_-=40$, respectively;  
(b) $\alpha_+=4$ and $\alpha_-=40$. 
}
\label{fig:profile}
\end{figure}
%%%%%%%%%%%%%%%%%%%%%%%%%%%%%%%%%%%%%%%%%%%%%%%%%%%%%%%%%%%%%%%%%%%%%%%%

Figure~\ref{fig:profile} shows that the exact solution [Eq.~(\ref{eq:sol1_1})] for $\alpha_+=\alpha_- \geq 1$ reproduces the numerical results with high accuracy.  
For $\alpha_+=\alpha_-\geq 1$, the approximate solution [Eq.~(\ref{eq:asymalphaneq1})] also closely follows the exact profile (not shown).  
When $\alpha_+\neq\alpha_- \geq 1$, the hypergeometric approximation [Eq.~(\ref{eq:hyper1})] accurately represents the full potential profile.  
In both cases, the exponential decay with a power-law correction [Eq.~(\ref{eq:asym3})] is valid only in the near-electrode region.  
Deviations from the exponential form become more pronounced as $|\alpha_\pm|$ increase, reflecting the enhanced influence of the temperature gradient, even when $\alpha_+ \neq \alpha_-$.  
Nevertheless, the near-electrode decay remains governed by the effective screening length $\lambda_{\rm eff}$ [Eq.~(\ref{eq:lmdeff})] for $\alpha_\pm \geq 1$.  
The exponential form provides an lower bound for the exact results when $\alpha_+=\alpha_- \geq 1$.

%%%%%%%%%%%%%%%%%%%%%%%%%%%%%%%%%%%%%%%%%%%%%%%%%%%%%%%%%%%%%%%%%%%%%%%
\begin{figure}[h]
\begin{center}
\includegraphics[width=1\textwidth]{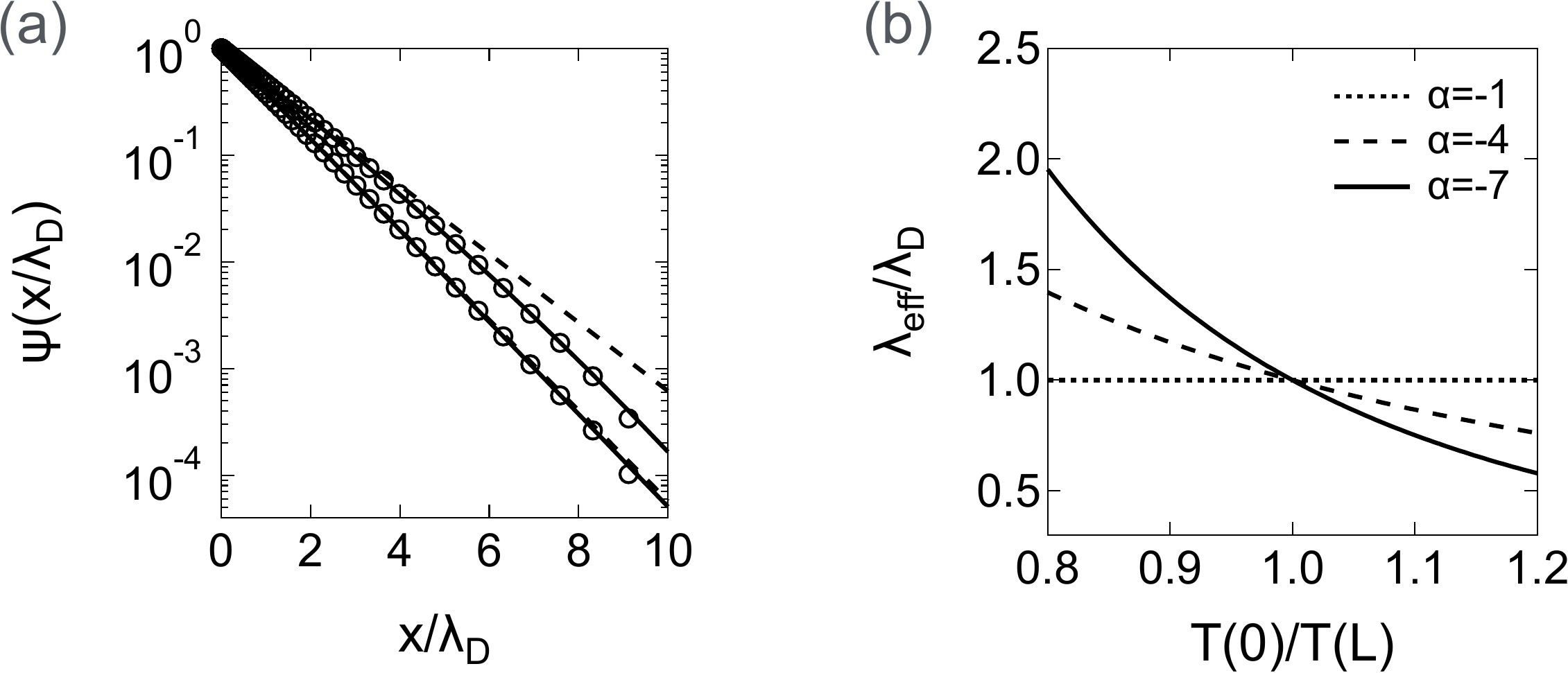}
\end{center}
\caption{Electrostatic potential as a function of the dimensionless distance from the electrode ($x/\lambda_D$) with $\psi(0)=1$, 
$T_{\rm r}=T(0)/T(L)=0.98$, and $g=0.00165$.  
Thick solid line: exact solution for $\alpha_+=\alpha_-$ [Eq.~(\ref{eq:sol2})];  
dashed line: exponential decay with a power-law correction [Eq.~(\ref{eq:asym3})] using $\lambda_{\rm eff}$ [Eq.~(\ref{eq:lmdeff})];  
circles: numerical solution of Eq.~(\ref{eq:DH}) with the additional boundary condition $\psi(10\lambda_D)=0$.  
(a) Upper and lower curves/circles correspond to $\alpha_+=\alpha_-=-4$ and $\alpha_+=\alpha_-=-40$, respectively;  
(b) $\lambda_{\rm eff}$ calculated from Eq.~(\ref{eq:lmdeff}). 
Short dashed, long dashed, and solid lines correspond to 
$\alpha_+=\alpha_-=-1$, $\alpha_+=\alpha_-=-4$, and $\alpha_+=\alpha_-=-7$, respectively.  
}
\label{fig:profile_alpham}
\end{figure}
%%%%%%%%%%%%%%%%%%%%%%%%%%%%%%%%%%%%%%%%%%%%%%%%%%%%%%%%%%%%%%%%%%%%%%%%

Figure~\ref{fig:profile_alpham} (a) shows that the exact analytical solution [Eq.~(\ref{eq:sol2})] for $\alpha_+=\alpha_- < 1$ reproduces the numerical results with high accuracy.  
For $\alpha_+=\alpha_-<1$, the asymptotic approximation [Eq.~(\ref{eq:asymalphaneq1})] also closely matches the exact profile (not shown).  
When $\alpha_+\neq\alpha_-<1$, the hypergeometric approximation [Eq.~(\ref{eq:hyper1})] accurately captures the full potential profile (not shown).  
The exponential decay with a power-law correction [Eq.~(\ref{eq:asym3})] remains valid only in the near-electrode region.  
Deviations from the exponential form become more pronounced as $\alpha_\pm$ increase, reflecting the enhanced influence of the temperature gradient.  
Nevertheless, the near-electrode decay is consistently governed by the effective screening length $\lambda_{\rm eff}$ [Eq.~(\ref{eq:lmdeff})] for $\alpha_\pm < 1$.  
The exponential form serves as an upper bound for the exact potential profile when $\alpha_+=\alpha_-<1$.

When $\alpha_+=\alpha_- <-1$, $\lambda_{\rm eff}$ [Eq.~(\ref{eq:lmdeff}]
decreases with increasing $T_{\rm r}$, corresponding to a lower electrode temperature, as shown in Fig.~\ref{fig:profile_alpham} (b).  
For $\alpha_+ < 0$ and $\alpha_- < 0$, ions tend to 
deplete on the cold side, increasing the screening length, 
whereas ions tend to  
accumulate on the hot side, decreasing the screening length.
Because ions naturally accumulate on the colder electrode side in the absence of the Soret effect, 
this effect sets in when $\alpha_+ = \alpha_- < -1$.

%Constant Potential (Dirichlet): $\psi(0)=V_0$ \\
%Constant Surface Charge (Neumann):  $d \psi/dx(0)=-\sigma/(\epsilon_r \epsilon_0)$ 

%%%%%%%%%%%%%%%%%%%%%%%%%%%%%%%%%%%%%%%
\section{Discussion}
%%%%%%%%%%%%%%%%%%%%%%%%%%%%%%%%%%%%%%%

We examine the potential of zero charge (PZC) within the non-isothermal Poisson--Boltzmann framework [Eq.~(\ref{eq:PB})], \cite{HORIKE_25} which remains nonlinear. \cite{Markovich_16}  
%%% added 
For symmetric thermodiffusion parameters ($\alpha = \alpha_+ = \alpha_-$),  we present the generalized Grahame equation near the PZC 
in Appendix~D. \cite{Grahame_47} 
%%% added

Combining Eqs.~(\ref{eq:APB5}) and (\ref{eq:BCsigma}) gives the $\sigma$--$\psi_0$ relation, 
\begin{align}
\frac{q^2}{4\epsilon_{\rm r}\epsilon_0k_{\rm B}T(L)}
\frac{\sigma^2}{q^2}
\approx
n_{\rm b}
\left(\frac{T(L)}{T_0}\right)^{\!\alpha-1}
\!\left[
\cosh\!\left(\frac{z_+q\psi_0}{k_{\rm B}T_0}\right)
-1
\right],
\label{eq:sigmasq}
\end{align}
where we have performed partial integration and assumed $\psi(L) = 0$, as in Eq.~(\ref{eq:gGC1}).  
The PZC corresponds to $\sigma = 0$ and exists irrespective of $T_0$, consistent with the isothermal case. \cite{Markovich_16}

The differential capacitance, $C_{\rm dc} = d\sigma/d\psi_0$, is given by
\begin{align}
C_{\rm dc} =
\frac{\epsilon_{\rm r}\epsilon_0}{\lambda_{\rm D}}
\left(\frac{T(L)}{T_0}\right)^{(\alpha+1)/2}
\cosh\!\left(\frac{z_+q\psi_0}{2k_{\rm B}T_0}\right).
\label{eq:capa1}
\end{align}
By substituting the effective screening length defined in Eq.~(\ref{eq:DYtempga2}), Eq.~(\ref{eq:capa1}) can be expressed in a compact, physically meaningful form:
\begin{align}
C_{\rm dc} =
\frac{\epsilon_{\rm r}\epsilon_0}{\lambda_{\rm eff}}
\cosh\!\left(\frac{z_+q\psi_0}{2k_{\rm B}T_0}\right).
\label{eq:capa1_1}
\end{align}
The double-layer capacitance is therefore characterized by the effective screening length $\lambda_{\rm eff}$.  
Equation~(\ref{eq:capa1_1}) shows that $C_{\rm dc}$ attains its minimum at the PZC, where the right-hand side of Eq.~(\ref{eq:sigmasq}) vanishes.  
The thermodiffusion parameter $\alpha$ can be estimated from Eq.~(\ref{eq:capa1}) by comparing $C_{\rm dc}$ measured under a temperature gradient with that of the isothermal electrical double layer.  
It is also known that $C_{\rm dc}$ can exhibit maxima when steric effects are included in the isothermal case,  
\cite{Bikerman_42,Freise_52,Borukhov_97,BORUKHOV_00,Kilic_07,Kornyshev_07,BAZANT_09,Adar_18}  
where the minimum of $C_{\rm dc}$ coincides with the PZC.  
In this study, we focus exclusively on the minimum, leaving the analysis of possible maxima, which require explicit treatment of steric effects, for future work.  
Such extensions may incorporate electrostatic interactions coupled with Soret and steric effects. \cite{deGroot_62,Kishikawa_10,Wurger_10,Odagiri_12,Penington_12}

Deviations from the classical Debye--H\"uckel screening length have been experimentally 
reported in ionic liquids and concentrated electrolytes even under isothermal 
conditions.\cite{Gebbie2013PNAS,Gebbie2015PNAS,Smith2016JPCL,Lee2017PRL,Gaddam2019Langmuir} 
According to large-scale molecular dynamics simulations, electrostatic screening lengths 
are not extremely anomalous but are still enhanced relative to classical predictions,\cite{Zeman2020ChemComm} 
which has motivated theoretical models that incorporate finite ion size, strong correlations, 
and clustering to rationalize these observations.\cite{Adar2019PRE,Haertel2023PRL,Dinpajooh2024JCP} 
Recent comparative reviews and perspective articles have highlighted open questions regarding 
the interpretation of, and apparent discrepancies between, experimentally determined screening 
lengths in concentrated electrolytes.\cite{Jaeger2023Faraday,Elliott2024CPL}
%%%%%%%%%%%%%%%%%%%%%%%%%%%%%%%%%%%%%%%

%%%%%%%%%%%%%%%%%%%%%%%%%%%%%%%%%%%%%%%

%%%%%%%%%%%%%%%%%%%%%%%%%%%%%%%%%%%%%%%
\section{Conclusion}
%%%%%%%%%%%%%%%%%%%%%%%%%%%%%%%%%%%%%%%
Previously, we theoretically evaluated various types of interfacial capacitance arising from 
asymmetric coverage of adsorbed ions on hot and cold electrodes as a possible origin of the 
large ionic Seebeck effect in ionic thermoelectric supercapacitors, and showed that the 
dominant contribution comes from the Stern-layer capacitance. Although numerical simulations 
indicated that the diffuse-layer capacitance is smaller than the Stern-layer capacitance, 
general theoretical expressions for the diffuse-layer capacitance and the associated screening 
length have not yet been derived.\cite{HORIKE_25} 

Here, we focus on the diffuse-layer capacitance under a temperature gradient in the limit of a 
small electrostatic potential difference. We introduce the Eastman entropy of transfer, 
$\hat{S}_\pm = \alpha_\pm k_{\rm B}$ for cations and anions, respectively, and analyze 
non-isothermal electric double layers in terms of the dimensionless Soret coefficients 
$\alpha_\pm$.

Analytical solutions of the generalized Debye--H\"uckel equation show that, under a temperature gradient, the electrostatic potential deviates from the classical exponential Debye--H\"uckel form.  
For identical thermodiffusion parameters ($\alpha_+=\alpha_-$), the potential is exactly expressed by a modified Bessel function 
[Eq.~(\ref{eq:sol1_1}) or (\ref{eq:sol2})], whereas the marginal case $\alpha=1$ exhibits an algebraic decay [Eq.~(\ref{eq:sol2_1})].  
Near the electrode, the effective screening length $\lambda_{\rm eff}$ can be defined 
[Eq.~(\ref{eq:lmdeff})]; $\lambda_{\rm eff}$ increases with temperature, indicating weaker screening on the hot side and stronger screening on the cold side due to thermally induced ion redistribution when $\alpha_\pm >-1$ and the opposite trends can be obtained for $\alpha_\pm <-1$.

Numerical calculations confirm that the exact solutions (modified Bessel or algebraic forms) reproduce the full potential profile for $\alpha_+=\alpha_-$, and that the hypergeometric approximation accurately describes the asymmetric case $\alpha_+\neq\alpha_-$.  
Although the exponential form with a power-law correction provides a useful near-electrode estimate, the overall potential remains non-exponential, particularly for large $\alpha_\pm$.  
These findings demonstrate that thermodiffusion substantially modifies the electric double-layer structure, with direct implications for ionic transport and electrostatic screening in non-isothermal electrolytes.

The non-isothermal Poisson--Boltzmann analysis demonstrates that the differential capacitance is governed by the thermodiffusion parameter $\alpha_\pm$ through its dependence on the effective screening length $\lambda_{\rm eff}$.  
The minimum of the differential capacitance coincides with the potential of zero charge (PZC) even in the presence of a temperature gradient.  
These findings provide a coherent theoretical framework for understanding the coupled effects of electrostatics and thermodiffusion in non-isothermal electrolyte systems.

The dependence of the effective screening length on the temperature difference between the cold 
and hot electrodes is found to be small for reasonable values of the dimensionless Soret coefficients 
and for temperature ranges relevant to ionic thermoelectric supercapacitors. 
This supports our previous conclusion that the Stern-layer capacitance, rather than the diffuse-layer 
capacitance, governs the large ionic Seebeck effect in ionic thermoelectric supercapacitors.\cite{HORIKE_25}

Recently, the voltage response under a suddenly applied temperature gradient has been investigated. \cite{Stout_17,Janssen_19,Sehnem_21,Liu_21}
In the limit of a weak temperature gradient, the voltage response was shown to depend on the Debye timescale, where $\lambda_{\rm D}$, rather than $\lambda_{\rm eff}$, 
appears because the Soret coefficients enter only through the boundary conditions. \cite{Stout_17,Janssen_19}
In this work, we restrict our analysis to steady-state conditions and do not consider the temporal evolution of the double-layer thickness induced by the Soret effect, 
which may be important for describing the voltage response beyond the weak temperature gradient limit.

%The generalized Grahame equation provides a relation between the electrode potential, surface charge density, and temperature distribution.  
%%%%%%%%%%%%%%%%%%%%%%%%%%%%%%%%%%%%%%%%%%%%%%%%%%%%%%%%%%%%%%%%%%%%%%
% ACKNOWLEDGMENTS %%%%%%%%%%%%%%%%%%%%%%%%%%%%%%%%%%%%%%%%%%%%%%%%%%%%
%%%%%%%%%%%%%%%%%%%%%%%%%%%%%%%%%%%%%%%%%%%%%%%%%%%%%%%%%%%%%%%%%%%%%%
%\newpage
%\noindent{\Large\bf Acknowledgment}
%\vspace{0.5cm}
\acknowledgments
%This work was supported by a Grant--in--Aid for Scientific Research (no. 20H02699) from the Ministry of Education, Culture, Sports, Science and Technology of Japan.

\section*{Data Availability Statement}
The data that support the findings of this study are available from the corresponding author upon reasonable request. 

\newpage 
\renewcommand{\theequation}{A.\arabic{equation}} 
\setcounter{equation}{0}  % reset counter     
\section*{Appendix A. Derivation of Eq. (\ref{eq:gGC1})}
%\section*{Appendix A. }

By linearizing Eq.~(\ref{eq:PB}), we obtain
\begin{multline}
\nabla^2 \psi(r) = -\frac{q}{\epsilon_{\rm r} \epsilon_0} 
\sum_{j} z_j n_j (L) 
\left(\frac{T(L)}{T(r)} \right)^{\alpha_j}
- \\
\frac{q^2}{\epsilon_{\rm r} \epsilon_0} 
\sum_{j} z_j^2 n_j (L) 
\left(\frac{T(L)}{T(r)} \right)^{\alpha_j} 
\int_r^L dr_1 \frac{1}{k_{\rm B} T(r_1)} \frac{d \psi}{d r_1} .
\label{eq:linearGC1}
\end{multline}
We impose the electroneutrality condition 
$\sum_{j} z_j q\, n_j(L) = 0$ at $r = L$, where $L$ is chosen to be much larger than the 
thickness of the electric double layer, but still smaller than the characteristic length 
scales of convection and deviations from the linear temperature gradient.
Under this charge neutrality condition, 
$\sum_{j} z_j n_j (L)=0$, 
Eq.~(\ref{eq:linearGC1}) reduces to
\begin{align}
\nabla^2 \psi(r) = -\frac{q^2}{\epsilon_{\rm r} \epsilon_0} 
\sum_{j} z_j^2 n_j (L)
\left(\frac{T(L)}{T(r)} \right)^{\alpha_j} 
\int_r^L dr_1 \frac{1}{k_{\rm B} T(r_1)} \frac{d \psi}{d r_1} .
\label{eq:linearGC2}
\end{align}
Differentiating both sides of Eq.~(\ref{eq:linearGC2}) with respect to $r$ yields
\begin{align}
\frac{d}{dr}\nabla^2 \psi(r)
= \frac{q^2}{\epsilon_{\rm r} \epsilon_0 k_{\rm B} T(L)} 
\sum_{j} z_j^2 n_j (L)
\left(\frac{T(L)}{T(r)} \right)^{1+\alpha_j} 
\frac{d \psi}{dr} .
\label{eq:linearGC3}
\end{align}
For a small temperature gradient, the linearized form becomes Eq. (\ref{eq:gGC1}), 
where we set $\psi(L)=0$, and $\psi(r)$ hereafter denotes $\psi(r)-\psi(L)$.

\newpage
\renewcommand{\theequation}{B.\arabic{equation}} 
\setcounter{equation}{0}  % reset counter     
\section*{Appendix B. Derivation of Eq. (\ref{eq:asym3})}
%\section*{Appendix A. }

An approximate solution of Eq.~(\ref{eq:DH1dim}) can be written as \cite{zaitsev02}
\begin{align}
\psi(x)=\psi_{\rm c}\, f(x)^{-1/4} 
\exp \left(-\frac{\kappa}{\sqrt{\sum_{j} z_j^2}} 
\int_0^x dx_1 \sqrt{f(x_1)} \right),
\label{eq:asym1}
\end{align}
where $f(x)$ is defined by
\begin{align}
f(x)=\sum_{j} z_j^2 
\left(\frac{T(L)}{T(x)}\right)^{1+\alpha_j} 
= \sum_{j} z_j^2 \left(\frac{1}{T_{\rm r}+g x}\right)^{1+\alpha_j}.
\label{eq:f1}
\end{align}

A further simplification follows from a Taylor expansion, 
\begin{align}
\int_0^x dx_1 \sqrt{f(x_1)} \approx \sqrt{f(0)}\,x + \cdots ,
\label{eq:dfx}
\end{align}
which reduces Eq.~(\ref{eq:asym1}) to
\begin{align}
\psi(x)\approx \psi_{\rm c}\, f(x)^{-1/4} 
\exp \left(-\frac{\kappa \sqrt{f(0)}}{\sqrt{\sum_{j} z_j^2}}\,x \right),
\label{eq:asym2}
\end{align}
where $\psi_{\rm c}$ is redefined to absorb constants.  
By substituting Eq.~(\ref{eq:f1}), we obtain Eq. (\ref{eq:asym3}), 
where an effective screening length under a temperature gradient is defined by Eq. (\ref{eq:lmdeff}).

When $\alpha_- \neq \alpha_+$ and $\alpha_\pm \neq 1$, 
a more accurate evaluation of Eq.~(\ref{eq:asym1}) compared to Eq.~(\ref{eq:asym3}) can be obtained using 
\begin{multline}
\int dx \,\sqrt{f(x)} 
= - \frac{2\left(T_{\rm r} + g x \right)^{(1-\alpha_-)/2}}{\left(\alpha_- -1\right)g} |z_-| \\
\times \, \mbox{}_2 F_1 \left(-\tfrac{1}{2}, \tfrac{-1+\alpha_-}{2(\alpha_+ - \alpha_-)};\, 
\tfrac{1+\alpha_- - 2\alpha_+}{2(\alpha_- - \alpha_+)};\, -\left(T_{\rm r}+g x\right)^{\alpha_- - \alpha_+} \right),
\label{eq:hyper1}
\end{multline}
where $z_+=z_-$ and $\mbox{}_2F_1(a,b;c;z)$ denotes the hypergeometric function \cite{abramowitzstegun}.  

\newpage
\renewcommand{\theequation}{C.\arabic{equation}} 
\setcounter{equation}{0}  % reset counter     
\section*{Appendix C. Derivation of Eq. (\ref{eq:sol1})}
%\section*{Appendix C. }

Eq.~(\ref{eq:gGCa}) can be simplified by introducing a new variable $z = T_{\rm r} + g x$,
\begin{align}
z^2 \frac{d^2 \psi(z)}{dz^2} = z^{\alpha-1} \frac{\kappa^2}{g^2} \psi(z).
\label{eq:Ap1}
\end{align}

We consider functions $\phi(z)$ and $\Phi(z)$ related by $\phi(z) = z^{c_1} \Phi(c_2 z^{c_3})$.  \cite{Beals_2010}
If $\Phi(z)$ satisfies the modified Bessel differential equation,
\begin{align}
z^2 \frac{d^2 \Phi(z)}{dz^2} + z \frac{d \Phi(z)}{dz} - (z^2 + \nu^2) \Phi(z) = 0,
\label{eq:Ap2}
\end{align}
then $\phi(z)$ satisfies modified Lommel's equation,
\begin{align}
z^2 \frac{d^2 \phi(z)}{dz^2} + (1 - 2 c_1) z \frac{d \phi(z)}{dz} - \left(c_2^2 c_3^2 z^{2c_3} - c_1^2 + c_3^2 \nu^2 \right) \phi(z) = 0.
\label{eq:Ap3}
\end{align}

By comparing Eq.~(\ref{eq:Ap1}) with Eq.~(\ref{eq:Ap3}) and assuming $\alpha \neq 1$, we identify
\begin{align}
c_1 = \frac{1}{2}, \quad c_3 = \frac{1-\alpha}{2}, \quad c_2 = \pm \frac{\kappa}{g c_3}, \quad \nu = \pm \frac{1}{1-\alpha}.
\end{align}

Therefore, the general solution of Eq.~(\ref{eq:Ap1}) can be expressed as a linear combination of
\begin{align}
z^{1/2} I_{\pm 1/(\alpha-1)}\!\left[\pm \frac{2 (\kappa/g) z^{(1-\alpha)/2}}{\alpha-1} \right] 
\quad \text{and} \quad
z^{1/2} K_{\pm 1/(\alpha-1)}\!\left[\pm \frac{2 (\kappa/g) z^{(1-\alpha)/2}}{\alpha-1} \right],
\end{align}
where the double signs are taken in the same order to ensure the electrostatic potential remains real.  
Transforming back to the original variable $x$ then yields Eq.~(\ref{eq:sol1}).

%%% added 
\newpage
\renewcommand{\theequation}{D.\arabic{equation}}  
\setcounter{equation}{0}  % reset counter     
\section*{Appendix D. Generalized Grahame Equation for Low Surface Charge Density}

For symmetric thermodiffusion parameters ($\alpha = \alpha_+ = \alpha_-$), Eq.~(\ref{eq:PB}) yields
\begin{align}
\frac{d}{dx}\!\left( \frac{d\psi}{dx} \right)^{\!2}
= 2 \frac{d\psi}{dx}
\!\left[
-\frac{q n_{\rm b}}{\epsilon_{\rm r}\epsilon_0}
\!\left(\frac{T(L)}{T(x)}\right)^{\!\alpha}
\sum_j z_j
\exp\!\left(
\!\int_x^L\! dx_1\,
\frac{z_j q}{k_{\rm B}T(x_1)}\frac{d\psi}{dx_1}
\right)
\right].
\label{eq:APB0}
\end{align}

Integrating gives \cite{HORIKE_25}
\begin{align}
\left( \frac{d\psi}{dx}\right)^2
= 2 \int_{x}^L dx' \frac{d\psi}{dx'}
\left[
\frac{q n_{\rm b}}{\epsilon_{\rm r} \epsilon_0}
\left(\frac{T(L)}{T(x')}\right)^\alpha  
\sum_{j} z_j 
\exp\!\left(\int_{x'}^L dx_1 \frac{z_j q}{k_{\rm B} T(x_1)} \frac{d \psi}{d x_1}\right)
\right],
\label{eq:APB1}
\end{align}
where the boundary condition at $x=L$ assumes charge neutrality in the bulk, 
\begin{align}
\left. \frac{d\psi}{dx} \right|_{x=L}=0.
\label{eq:APBBC1}
\end{align}

Since
\begin{align}
\frac{d}{dx} \exp \!\left(\int_x^L dx_1 \frac{z_j q}{k_{\rm B} T(x_1)} \frac{d \psi}{d x_1}\right)
= -\frac{z_j q}{k_{\rm B} T(x)} \frac{d \psi}{d x}
\exp \!\left(\int_x^L dx_1 \frac{z_j q}{k_{\rm B} T(x_1)} \frac{d \psi}{d x_1}\right),
\label{eq:APB2}
\end{align}
Eq.~(\ref{eq:APB1}) can be rewritten as
\begin{align}
\left( \frac{d\psi}{dx}\right)^2
= -2 \int_x^L dx' 
\left[
\frac{n_{\rm b} k_{\rm B}}{\epsilon_{\rm r} \epsilon_0} 
\frac{T(L)^\alpha}{T(x')^{\alpha-1}}
\sum_{j} 
\frac{d}{dx'} 
\exp \!\left(\int_{x'}^L dx_1 \frac{z_j q}{k_{\rm B} T(x_1)} \frac{d \psi}{d x_1}\right)
\right].
\label{eq:APB3}
\end{align}

By performing partial integration, Eq.~(\ref{eq:APB3}) becomes
\begin{multline}
\left( \frac{d\psi}{dx}\right)^2
= 2
\frac{n_{\rm b}k_{\rm B} T(L)}{\epsilon_{\rm r} \epsilon_0}
\Bigg[
\left(\frac{T(L)}{T(x)} \right)^{\alpha-1}
\sum_{j} 
\exp \!\left(\int_x^L dx_1 \frac{z_j q}{k_{\rm B} T(x_1)} \frac{d \psi}{d x_1}\right)
- 2 \\
- (1-\alpha) 
\int_x^L dx' 
\frac{T(L)^{\alpha-1}}{T(x')^{\alpha}}
\frac{d T(x')}{dx'}
\sum_{j} 
\exp \!\left(\int_{x'}^L dx_1 \frac{z_j q}{k_{\rm B} T(x_1)} \frac{d \psi}{d x_1}\right)
\Bigg].
\label{eq:APB4}
\end{multline}

Applying Gauss’s law at the electrode gives
\begin{align}
- \left. \epsilon_{\rm r}\epsilon_0\frac{d\psi}{dx}\right|_{x=0}=\sigma,
\label{eq:BCsigma}
\end{align}
where $\sigma$ is the effective surface charge density.  
The PZC corresponds to $\sigma = 0$.  
Accordingly, we introduce the approximation
\begin{align}
\exp \!\left(\int_{x'}^L dx_1 \frac{z_j q}{k_{\rm B} T(x_1)} \frac{d \psi}{d x_1}\right) \approx 1.
\label{eq:apprAPB4}
\end{align}
In previous analyses, we considered the case of high surface charge density and assumed the temperature gradient to be the smallest parameter, which led to slightly different final expressions.  
Substituting Eq.~(\ref{eq:apprAPB4}) into Eq.~(\ref{eq:APB4}) yields
\begin{align}
\left( \frac{d\psi}{dx} \right)^{\!2}
\approx
\frac{4n_{\rm b}k_{\rm B}T(L)}{\epsilon_{\rm r}\epsilon_0}
\left(\frac{T(L)}{T(x)}\right)^{\!\alpha-1}
\!\left[
\cosh\!\left(
\int_x^L dx_1\,
\frac{z_+ q}{k_{\rm B}T(x_1)}\frac{d\psi}{dx_1}
\right)
-1
\right].
\label{eq:APB5}
\end{align}

Now, we present the generalized Grahame equation near the PZC, \cite{Grahame_47} obtained by applying the inverse transformation of Eq. (\ref{eq:APB5}).
For low effective surface charge densities, the generalized Grahame equation follows from Eq.~(\ref{eq:sigmasq}) with the boundary condition Eq.~(\ref{eq:BCsigma}):  
\begin{align}
\left| \frac{z_+ q \psi_0}{k_{\rm B} T_0} \right|
= \cosh^{-1} \Biggl[
1 + \frac{\sigma^2}{4 \epsilon_{\rm r} \epsilon_0 n_{\rm b} k_{\rm B} T(L)}
\left(\frac{T_0}{T(L)}\right)^{\alpha-1} 
\Biggr].
\label{eq:APB10}
\end{align}
We adopt the standard convention $\cosh^{-1}(x) = \ln(x + \sqrt{x^2 - 1})$ for $x \ge 0$ \cite{NIST}.  
Equation~(\ref{eq:APB10}) can equivalently be expressed as
\begin{align}
\frac{z_+ q \psi_0}{k_{\rm B} T_0}
\approx
2 \sinh^{-1} \Biggl[
\frac{\sigma}{\sqrt{8 \epsilon_{\rm r} \epsilon_0 n_{\rm b} k_{\rm B} T(L)}}
\left(\frac{T_0}{T(L)}\right)^{(\alpha-1)/2}
\Biggr].
\label{eq:APB8}
\end{align}

%%%%%%%%%%%%%%%%%%%%%%%%%%%%%%%%%%%%%%%%%%%%%%%%%%%%%%%%%%%%%%%%%%%%%%
% REFERENCES %%%%%%%%%%%%%%%%%%%%%%%%%%%%%%%%%%%%%%%%%%%%%%%%%%%%%%%%%
%%%%%%%%%%%%%%%%%%%%%%%%%%%%%%%%%%%%%%%%%%%%%%%%%%%%%%%%%%%%%%%%%%%%%%
%\newpage
%\begin{thebibliography}{99}
%\vspace{0.5cm}
%
%\end{references}

\end{document}